# Small RNAs Establish Delays and Temporal Thresholds in Gene Expression

Stefan Legewie[1*], Dennis Dienst[2], Annegret Wilde[2], Hanspeter Herzel[1] and Ilka M. Axmann[1,2]

[1]Humboldt University Berlin, Institute for Theoretical Biology, Invalidenstrasse 43, D-10115 Berlin, Germany;
[2]Humboldt University Berlin, Institute of Biology, Chausseestrasse 117, D-10115 Berlin, Germany;

[*] To whom correspondence should be addressed. E-mail: s.legewie@biologie.hu-berlin.de, Telephone: +49 30 2093 9121

**Running title: Regulation by small RNAs.**

Non-coding RNAs are crucial regulators of gene expression in prokaryotes and eukaryotes, but it remains poorly understood how they affect the dynamics of transcriptional networks. We analyzed the temporal characteristics of the cyanobacterial iron stress response by mathematical modeling and quantitative experimental analyses, and focused on the role of a recently discovered small non-coding RNA, IsrR. We found that IsrR is responsible for a pronounced delay in the accumulation of *isiA* mRNA encoding the late-phase stress protein, IsiA, and that it ensures a rapid decline in *isiA* levels once external stress triggers are removed. These kinetic properties allow the system to selectively respond to sustained (as opposed to transient) stimuli, and thus establish a temporal threshold, which prevents energetically costly IsiA accumulation under short-term stress conditions. Biological information is frequently encoded in the quantitative aspects of intracellular signals (e.g., amplitude and duration). Our simulations reveal that competitive inhibition and regulated degradation allow intracellular regulatory networks to efficiently discriminate between transient and sustained inputs.

# Introduction

Non-protein-coding RNA regulators such as microRNAs (miRNA) and short interfering RNAs (siRNA) control diverse processes in metazoa including development, cell differentiation, and cell proliferation (1, 2). Recent research revealed that small non-coding RNAs (sRNAs) also play important roles in bacteria (3, 4), where they are mainly involved in the modulation of stress responses. Approximately 80 sRNAs have been identified in *E. coli*, many of which are evolutionary conserved (5). sRNAs are typically less than 200 nucleotides in size and are either encoded in *cis* or in *trans*. *Cis*-encoded sRNAs are transcribed from the antisense strand of their target mRNAs, and are thus perfectly complementary, while *trans*-encoded sRNAs have only limited complementarity (4). Most sRNAs inhibit gene expression employing a non-catalytic mechanism of action: Base-pairing with target mRNAs either interferes with ribosome binding and thus with target mRNA translation, or even induces degradation of the whole sRNA-target complex (4, 6).

Previous mathematical modeling indicated that sRNA-mediated regulation may allow for faster control over gene expression than transcriptional and post-translational regulation (7). Moreover, it was shown theoretically and experimentally that small non-coding RNAs efficiently suppress steady state target mRNA accumulation if the mRNA transcription rate is low, while they have little impact at higher mRNA transcription rates (8, 9). Levine et al (2007) argue that this phenomenon, termed a "threshold-linear response", efficiently prevents costly and potentially harmful expression of bacterial stress proteins under normal conditions.

In this study, we set out to analyze the impact of sRNAs on the temporal regulation of gene expression. Theoretical predictions derived from mathematical modeling are confirmed by quantitative experimental analyses of the iron stress response in a cyanobacterial organism (*Synechocystis* sp. PCC 6803). The IsiA (iron stress induced protein A) stress response protein, which is transcriptionally induced upon iron depletion or oxidative stress (10) is controlled by a naturally occurring antisense sRNA, IsrR (iron stress repressed RNA) (11). By comparing strains expressing different levels of IsrR sRNA, we find that IsrR is responsible for a pronounced delay in *isiA* induction. This delay ensures that iron stress proteins are expressed in a temporally ordered manner, with the "emergency" protein IsiA accumulating only if the stress duration exceeds a critical temporal threshold. Moreover, we find that IsrR sRNA ensures a rapid decline in *isiA* levels once external stress triggers are removed. IsiA expression must be tightly controlled, as it reduces photosynthesis efficiency in unstressed cells and becomes highly abundant under stress conditions. Our results reveal how the IsrR sRNA ensures that *isiA* accumulation is restricted to severe, prolonged and ongoing stress conditions.

## Results and Discussion

*Mathematical Model*

We implemented a mathematical model of sRNA-mediated regulation (schematically depicted in Fig. 1A), which includes synthesis and degradation of target mRNA and sRNA, respectively. Additionally, we considered a reversible association reaction between the target mRNA and the sRNA, and also took degradation of the resulting heteroduplex ('Pair') into account. In contrast to published models (7, 8), the heteroduplex concentration was treated as a dynamic variable, mainly because our experimental analyses of the cyanobacterial iron stress response did not allow to distinguish between free and sRNA-bound *isiA* mRNA. Previous studies indicated that heteroduplex association and dissociation proceed with rapid kinetics when compared to protein synthesis and degradation (12, 13). We therefore applied a rapid-equlibrium assumption so that the model depicted in Fig. 1A reduces to a two-variable system (see also the Supplement):

$$d[T_{tot}]/dt = v_{syn,T} - k_{deg,T} \cdot ([T_{tot}]-[Pair]) - k_{deg,P} \cdot [Pair]$$
$$d[S_{tot}]/dt = v_{syn,S} - k_{deg,S} \cdot ([S_{tot}]-[Pair]) - k_{deg,P} \cdot [Pair] \quad (1)$$

Here, $[T_{tot}]$ and $[S_{tot}]$ indicate the dynamical variables of the system which represent the total intracellular concentrations of the target mRNA and sRNA, respectively. More specifically, $[T_{tot}]$ equals the sum of the concentrations of free and sRNA-bound target mRNA (i.e., $[T_{tot}]$ = [Target] + [Pair]), and $[S_{tot}]$ = [sRNA] + [Pair] is defined similarily. Owing to the rapid equilibrium assumption, the concentration of the heteroduplex (i.e., [Pair]) is given by (see Supplement):

$$[Pair]=1/2 \cdot \left( [T_{tot}]+[S_{tot}]+K_{d,P} - \sqrt{\left([T_{tot}]+[S_{tot}]+K_{d,P}\right)^2 - 4 \cdot [T_{tot}] \cdot [S_{tot}]} \right) \quad (2)$$

Equations (1) and (2) constitute a reduced differential equation system that depends on the dynamical variables $[T_{tot}]$ and $[S_{tot}]$ only. From the total RNA concentrations, $[T_{tot}]$ and $[S_{tot}]$, one can calculate back to the individual concentrations (i.e., [Target], [sRNA] and [Pair]) by using Eq. 2 and the mass conservation definitions of $[S_{tot}]$ and $[T_{tot}]$. Numerical simulations were generally done using the units nM for intracellular concentrations and h for time. In Figs. 1 and 2, the concentrations on the y-axes are given in arbitrary units (A.U.), because no absolute experimental quantification of RNA expression was available.

The cyanobacterial stress response protein IsiA is induced by reactive oxygen species (ROS) and by iron depletion, and its expression is further modulated by a small non-coding RNA, IsrR (Fig. 1A). We measured the kinetics of *isiA* regulation to confirm the model predictions (see below). The input functions in the model were therefore chosen such that they reflect those in the cyanobacterial stress response. IsiA expression is known to be transcriptionally regulated by the iron-sensitive Fur (ferric uptake regulator) repressor, while the activity of the promoter controlling the antagonizing sRNA, IsrR, is not affected by cellular stress (Wolfgang Hess, pers. comm.). The synthesis rate of target mRNA ($v_{syn,T}$) is therefore modeled to be controlled by external stimuli, as indicated in Fig. 1A. For simplicity, it is assumed that the regulation of the *isiA* promoter activity occurs on a much faster time scale than downstream RNA synthesis and degradation. Thus, extracellular stimulation was simulated by a step-like change of $v_{syn,T}$ from one value to another. The Michaelis-Menten equation ($v_{syn,T} = V_{max,synT} \cdot$ Stimulus / ($K_{M,synT}$ + Stimulus)) was employed to simulate dose-response behavior (Fig. 1B).

*Theoretical Analysis of Steady State and Dynamical Behavior*

We systematically analyzed the steady state and dynamical behavior of the sRNA circuit by numerical simulations. In the following, we mainly focus on the scenario, where the heteroduplex is much less stable than the monomeric sRNA and target species, as such behavior was recently reported for the cyanobacterial stress response (11). However, the main conclusions drawn in this paper remain valid even if the sRNA merely acts as a competitive inhibitor of translation, which does not enhance target degradation (see 'Concluding Remarks')

Recent work by Levine et al. (8) revealed that sRNAs establish sharp thresholds in the steady state stimulus-response behavior of gene regulatory circuits. Similarly, our simulations also yielded an all-or-none expression pattern of target mRNA if physiologically relevant kinetic parameters were used (Fig. 1B, top, solid line). As discussed previously (8), sRNA-mediated regulation efficiently suppresses mRNA accumulation as long as the sRNA synthesis rate exceeds that of the target mRNA (i.e., if $v_{syn,T} < v_{syn,S}$). In Fig. 1B, we have $v_{syn,T} = v_{syn,S}$ if the stimulus level equals unity, and thus observe a threshold for stimulus = 1. As expected, the threshold disappeared in the absence of sRNA (dashed line) or in case the affinity of the heteroduplex was too low (not shown). Moreover, we found that sharp thresholds were typically accompanied by a mutually exclusive expression pattern of target mRNA and sRNA (solid and dashed-dotted lines in Fig. 1B). Remarkably, our previous experimental work revealed that the *isiA* mRNA and its modulator, the IsrR sRNA, accumulate in a mutually exclusive manner under various stimulation conditions (11). In further experiments, we analyzed *isiA* and IsrR amounts for varying iron concentrations in the medium (Supplemental Figure S2), and found that IsrR sRNA starts to accumulate only if *isiA* mRNA falls below a certain level. Finally, a comparison of wild-type cells with IsrR knockdown cells reveals that IsrR completely suppresses residual *isiA* levels under unstressed conditions (compare circle and square at t = 0 in Fig. 2A). Taken together, these data strongly suggest that a sharp threshold exists in the cyanobacterial iron stress response, and that the system operates near this threshold under physiological conditions.

We investigated the temporal dynamics around the threshold in order to derive experimentally testable predictions from the model. We focused on the time required to pass from one steady state to another ('response time'). Step-like increases in the external stimulus were applied, and the response time $t_{50}$ required to reach 50% of the difference between old and new steady state was calculated for the target mRNA. The dotted line in Fig 1B (bottom) reflects the response time of a reference system without sRNA, and is valid for both input scenarios considered in our analysis: (i) a step-like stimulus increase from no stimulation to the value indicated on the x-axis ('upregulation'); (ii) a stimulus decrease from the value indicated on the x-axis to no stimulation ('downregulation'). Our simulations of the sRNA circuit revealed that sRNAs speed up the upregulation kinetics for subthreshold stimuli, while they establish a delay upon suprathreshold stimulation (solid line). Moreover, we found that sRNAs always accelerate target downregulation even for suprathreshold stimuli (dashed line).

In the subthreshold regime the sRNA is present in excess over the target mRNA. Under these conditions, the sRNA can be assumed to be constant, so that mRNA degradation via the Pair intermediate behaves like a first-order decay reaction, which dominates over the much slower direct mRNA degradation step (see Supplement). The response times of mRNA up- and downregulation are known to be solely determined by the (fastest) first-order decay rate (14), which explains why sRNA-mediated regulation accelerates target regulation in both directions under subthreshold stimulation conditions. The dynamic phenomena we observed upon strong stimulation originate from nonlinear threshold effects in the sRNA circuit. The upregulation time lag arises, because the residual pool of free small RNAs needs to be cleared before the target mRNA can accumulate. Small RNAs accelerate downregulation, since the mRNA concentration quickly falls into a subthreshold range, where sRNA-mediated regulation efficiently degrades residual mRNA. This results in an abrupt termination of the mRNA expression time course, as can be seen from solid line in Fig. 2B. Moreover, the initial phase of mRNA down-regulation is accelerated in the presence of small non-coding RNAs (compare solid and dotted lines in Fig. 2B), as the pair intermediate is turned over at a higher rate than the free mRNA. Taken together, our simulations suggest that IsrR sRNA delays *isiA* induction upon stress, but speeds up *isiA* downregulation when the upstream trigger is removed.

*Experimental Verification of the Simulated Dynamic Behavior*

Quantitative experimental analyses were performed to confirm the simulations in Fig. 1B. The impact of sRNA-mediated regulation was investigated by comparing the kinetics of *isiA* mRNA accumulation in mutant strains expressing different levels of IsrR sRNA. The data points shown in Fig. 2A are densitometric quantifications of previously published measurements of *isiA* induction in response to hydrogen peroxide stress (11). These data revealed that the delay in *isiA* accumulation seen in wild-type cells can be further enhanced in cells overexpressing IsrR, while it is abolished in IsrR-depleted cells. The measured time courses thus agree well with the simulation result that sRNAs decelerate target mRNA induction.

We also measured the time course of *isiA* target mRNA downregulation after removal of the stress trigger (see Fig. S1A in the Supplement). Hydrogen peroxide could not be used as the stimulus in these experiments, since it induces irreversible oxidation of cellular components so that stress might actually persist even after hydrogen peroxide is removed. *isiA* accumulation was therefore induced by iron depletion (48 h), and *isiA* expression was subsequently downregulated by iron re-addition (t = 0 in Fig. 2B). The time course measurements after relief from iron stress were in accordance with the model predictions, as the decline in *isiA* levels was faster in wild-type cells when compared to knockdown cells harboring reduced levels of IsrR sRNA (Fig. 2B).

Having established a qualitative agreement between experiments and simulations, we next asked whether our model could also quantitatively reproduce the dynamics of *isiA* target mRNA up- and downregulation. The model parameters were estimated from the time course data shown in Figs. 2A and B. The trajectories of the best-fit model (solid lines in Figs. 2A and B) indicate that all measurements can be accurately described by a single set of kinetic constants. Remarkably, the best-fit parameters, summarized in Table 1, suggest that the heteroduplex is rapidly degraded, while the single stranded forms of *isiA* and IsrR are predicted to be much more stable than typical bacterial RNAs (15). A long half-life of *isiA* directly follows from the slow *isiA* downregulation kinetics in sRNA-depleted cells (Fig. 2B; blue line), while the stability of IsrR cannot be as straight-forwardly deduced from the time course data. We therefore directly measured the half-life of IsrR sRNA in unstressed cells, which express negligible amounts of *isiA* target mRNA (11). Very little degradation of IsrR occured within a 45 min time interval after incubation of cells with the general transcription inhibitor Rifampicin (see Fig. S1B in the Supplement). These data confirm that IsrR sRNA is stable and thus support our best-fit model.

*Pulse Filtering Properties of sRNA Circuits*

The above RNA measurements (Figs. 2A and B) did not discriminate between IsrR-bound ('pair') and free ('target') *isiA* RNA. In order to get more direct insights into the regulation of *isiA* action by IsrR sRNA, we simulated time courses of *free* (i.e., biologically active) *isiA* target mRNA using the best-fit parameters. Figure 2C shows model responses to a step-like suprathreshold pulse stimulation, and reveals that regulation by small RNAs gives rise to a sharp, spike-like time course (solid line), when compared to a system devoid of sRNA (dashed line). More specifically, target mRNA accumulation is completely suppressed until all sRNA is degraded (indicated by black circle), and the mRNA decline terminates abruptly, once the sRNA (re-)starts to accumulate.

Taken together, our simulations indicate that IsrR-mediated control serves to prevent premature and unnecessarily prolonged *isiA* synthesis. This is consistent with the hypothesis that IsiA establishes a second line of defence against iron depletion, and with the fact that its expression occurs relatively late during iron stress (16). The delay established by sRNA-mediated regulation thus enables cells to induce both early- and late-phase stress proteins in response to a single stress trigger in a temporally ordered manner. Two lines of evidence further suggest that spike-like *isiA* expression (Fig. 2C) is beneficial to the cellular energy budget. First, *isiA* becomes the most abundant transcript in cells subjected to oxidative stress (17). Second, IsiA expression saves photosynthesis in stressed cells, but decreases photosynthesis efficiency and thus energy production in non-stressed cells (16). Our analyses indicate that IsrR sRNA-mediated control allows avoiding lavish *isiA* accumulation unless cells are subjected to severe and prolonged stress.

We sought to further confirm our hypothesis that sRNA circuits suppress short stimuli, but efficiently transmit prolonged inputs. The time courses of target mRNA expression were, therefore, simulated for step-like stimulus pulses, and the time course maximum was analyzed as a function of pulse duration (Fig. 2D). We found that the best-fit model (Table 1) indeed responds to the pulse duration in a highly ultrasensitive manner (Hill coefficient $n_H \approx 3.5$), while a much more gradual duration response is seen for a system without sRNA. A similar result was obtained when we analyzed the integral under the target mRNA time courses (Fig. 2D, inset), thus further confirming that sRNAs establish pulse filtering and temporal thresholds in biochemical signaling networks.

*Concluding Remarks*

Using a combination of mathematical modeling and quantitative experimental analyses, we have shown that sRNAs establish delays and (steady state and temporal) thresholds in gene expression. In order to allow for a comparison of our simulations with experimental data, we analyzed the dynamical behavior of models with different sRNA expression levels for a given stimulus level, and in most cases compared the wild-type model with a model devoid of sRNA. It should be noted that, from a systems theoretical point of view, the dynamical behavior of both models (+/- sRNA) is not directly comparable for a given stimulus level, as they differ in their steady state dose-response curves. However, it can be seen in Fig. 1B that sRNA-mediated regulation still establishes a delay in mRNA upregulation and accelerates mRNA downregulation if both models (+/- sRNA) are compared for a given suprathreshold steady state activation level. In this context, it is important to note that the up- and downregulation response times of the model without sRNA are stimulus-invariant (dashed horizontal line in Fig. 1B, bottom).

The key finding of our paper is that sRNA-mediated regulation establishes a sign-sensitive delay for supra-threshold stimulation (Figs. 2A and B), and this conclusion does not depend on the precise dose-response behavior or on the kinetic parameters chosen. Here, the term 'sign-sensitive delay' denotes that a delay is observed exclusively for mRNA upregulation, but not for mRNA downregulation (18). This sign-sensitive delay is responsible for the pulse-filtering behavior shown in Figs. 2C and D: The delay in mRNA upregulation helps to filter out short supra-threshold transients. However, *ultrasensitive* pulse-filtering with a Hill coefficient significantly larger than unity additionally requires that no such delay and in particular no memory effects arise if the stimulus is removed. The experimental data presented in the paper rules out that mRNA downregulation occurs with a delay (Fig. 2B), and thus strongly supports that ultrasensitive pulse-filtering occurs in the cyanobacterial iron stress response. However, the mRNA downregulation kinetics might depend on the history of the system (e.g., short vs. long mRNA upregulation), so that explicit pulse-stimulation experiments are required to further prove the existence of ultrasensitive pulse-filtering.

In our simulations, we mainly focused on the scenario, where the heteroduplex is less stable than the single stranded sRNA and target species. However, several bacterial sRNAs do not enhance target degradation, but merely act as competitive inhibitors of translation (6). Importantly, small RNAs acting in this way still establish sharp thresholds and delays, as translation of subthreshold target mRNA levels is efficiently suppressed by sequestration into the heteroduplex (19) (Supplement). Eukaryotic miRNA action can also be described by the model scheme depicted in Fig. 1A, as miRNAs either competitively inhibit translation or induce mRNA degradation. However, miRNAs frequently remain intact after target degradation, and can guide the recognition and destruction of additional messages (1). Our numerical analysis revealed that the kinetic properties described above remain valid even if a fraction of the sRNA remains intact during the pair degradation reaction (Supplement) (8). Thus, the main results of this paper apply for eukaryotic systems as well, although this remains to be confirmed in more detailed models of miRNA action (20).

In Figs. 1 and 2, we analyzed the dynamic characteristics of mRNA expression. However, the kinetic properties of the sRNA circuit may be obscured by a slow response at the level of protein expression. We therefore investigated numerically whether our conclusions regarding pulse-filtering continue to hold at the protein level. The simulation results (Fig. S5) demonstrate that for the best-fit parameters pulse-filtering is preserved at the protein level even if IsiA protein is assumed to be relatively stable, with a half-life of 10 h. Up to now, studies focusing on the stability of IsiA protein are missing in the literature. However, half-life measurements of the CP43 photosynthesis protein homologous to IsiA revealed a half-life of about an hour under stress conditions (21), and a similar rapid turnover was also reported for another photosynthesis protein, D1 (22). These data suggest that IsiA protein is short-lived in the experimental setup we have chosen, and that the pulse filtering property discussed here for the RNA level is observed at the level of proteins as well.

Cells face a specificity problem, as broadly overlapping signaling pathways are activated by diverse stimuli. Biological information is therefore encoded in the quantitative aspects of the signal, such as amplitude and duration (23, 24). An important role for the signal duration in the initiation of cell fate decisions was described for various biological networks including MAPK signaling (23), TGFβ signaling (25), cAMP signaling (26), and NF-κB signaling (27). Previous work indicated that multistep regulation in the form of feedforward loops (28) and multisite phosphorylation (29, 30) allows cells to discriminate transient and sustained stimuli. In this paper, we identified competitive inhibition and/or regulated degradation as alternative plausible mechanisms for duration decoding (Fig. 2D). We propose that functional analysis of small RNAs might explain why some genes are selectively transcribed upon sufficiently long stimulation (27, 31). The results are likely to be of broader relevance, because regulation by protein-protein interactions frequently involves competitive inhibition and regulated degradation as well (32-35).

## Materials and Methods

*Mathematical Modeling*: Numerical simulations were done in Matlab 7.3 (codes available upon request). PottersWheel, a multi-experiment fitting toolbox for Matlab programmed by Thomas Maiwald (www.potterswheel.de), was used for parameter estimation in Figs. 2A and B. The Hill coefficient $n_H$ of the red curve in Fig. 2D was estimated by using the formula $n_H = \log(81) / \log(D_{90}/D_{10})$, where $D_{90}$ and $D_{10}$ are the stimulus durations required for the target amplitude to reach 90% and 10% of the steady state, respectively (36).

*Experimental Part*: Bacterial strains and standard growth conditions were as described (11) with minor modifications (30°C; 50 µmol photons $m^{-2}$ $s^{-1}$). Exponentially growing *Synechocystis sp.* PCC 6803 liquid cultures ($OD_{750}$ 0.7) were washed four times with iron-free medium and further grown for 48 hours in fermenters supplied with a continuous stream of air. Iron pulse was achieved by the addition of 43 nmol ferric ammonium citrate per 100 mL culture at an $OD_{750}$ of 0.8. RNA isolation and analysis procedures were carried out as described before (11). RNA stability was determined as described (3) with the exception that after the addition of rifampicin arresting transcription cells were harvested on ice at different time points, followed by short-term centrifugation and resuspension in TRIzol reagent (Invitrogen). Total RNA was isolated with TRIzol reagent (Invitrogen) and separated on 10% polyacrylamide-urea or 1.3% agarose formaldehyde gels followed by Northern blotting. After hybridization with radio-labeled probes against either *isiA* or IsrR (16S or 5S rRNA as standard), Northern blot signals were detected and analyzed on a Personal Molecular Imager FX system with QUANTITY ONE software (Bio-Rad). For loading control of Northern blots, we used hybridization signals of rRNA as an internal standard.

# Tables

**Table 1**: Kinetic parameters of the best-fit model.

| Figure | Fig. 2A | Fig. 2B |
|---|---|---|
| $v_{syn,T}$ [nM h$^{-1}$] | 247.2 (basal level) <br> 1136 (after stimulus increase) | 996.9 (basal level) <br> 41.9 (after stimulus removal) |
| $k_{deg,T}$ [h$^{-1}$] | 0.42 | 0.42 |
| $v_{syn,S}$ [nM h$^{-1}$] | 516.6 (red line) <br> 706.9 (green line) <br> 145.3 (blue line) | 516.6 (red line) <br> 145.3 (blue line) |
| $k_{deg,S}$ [h$^{-1}$] | 0.35 | 0.35 |
| $K_{D,P}$ [nM] | 0.0045 | 0.0045 |
| $k_{deg,P}$ [h$^{-1}$] | 13.75 | 13.75 |

# Figure Legends

**Figure 1**: Theoretical analysis of gene expression regulation by small RNAs. **(A)** Schematic representation of the mathematical model. The target mRNA associates with a small non-coding RNA ('sRNA') to form a heteroduplex ('pair'). Molecular species and events relevant for the cyanobacterial stress response (ROS = reactive oxygen species; iron depletion; *isiA* and IsrR) and kinetic parameters are indicated in grey. **(B)** Steady state (top) and dynamical (bottom) response to varying stimulus strength. The top graph indicates the steady state levels of (free) sRNA (dashed-dotted line) and target mRNA (solid line), and reveals a mutually exclusive expression pattern characterized by a sharp threshold. For comparison, the gradual target mRNA dose-response curve in the absence of sRNA expression is also shown (dotted line). The bottom graph shows the response time of total target expression (i.e., sum of target and pair) as a function of the stimulus level. The solid line ('upregulation') depicts the response time $t_{50}$ required to reach 50% of the difference between old and new steady state upon a step-like increase in the stimulus level from no stimulation to the level indicated on the x-axis. Similarly, $t_{50}$ was also calculated for a step-like drop in the stimulus level from the value on the x-axis to no stimulation ('downregulation'). The dotted line depicts the response time of a system without sRNA and is valid for both up- and downregulation. See Supplemental Table S1 for kinetic parameters.

**Figure 2**: Experimental verification of simulated dynamical behaviour (A and B) and physiological relevance (C and D). **(A)** Regulation by small RNAs establishes a delay in target mRNA accumulation. The time course of *isiA* target mRNA accumulation in response to $H_2O_2$ treatment was measured in wild-type cells (circles), in cells harbouring reduced amounts of IsrR sRNA (squares), and in cells overexpressing IsrR sRNA (crosses). The measurements agree well with the corresponding simulations (lines; see Table 1 for kinetic parameters). **(B)** The decline in target mRNA (*isiA*) levels is faster in wild-type cells (circles) when compared to cells depleted of IsrR sRNA (squares), as expected from corresponding simulations (lines; see Table 1 for kinetic parameters). Accumulation of *isiA* target mRNA was induced by iron starvation (48h), and then *isiA* expression was blocked by iron re-addition at t = 0 (see text). **(C)** Regulation by small RNAs gives rise to a sharp, spike-like time course of *free* target mRNA in response to step-like pulse stimulation (solid line), when compared to a system without sRNA (dashed line). In particular, target mRNA accumulation is completely suppressed until all sRNA is degraded (indicated by black circle). **(D)** Small RNAs establish pulse filtering in gene expression. Target mRNA accumulation in response to step-like stimuli of different duration (but of constant strength) was simulated, and the time course maximum or the integral similar to the hatched area in C (inset) is plotted as a function of pulse length. Highly ultrasensitive pulse duration decoding (Hill coefficient $n_H \approx 3.5$) is seen in the presence (solid line), but not in the absence (dashed line), of sRNA-mediated regulation. See Supplemental Table S1 for kinetic parameters.

# Figures

**Figure 1**: Theoretical analysis of gene expression regulation by small RNAs.

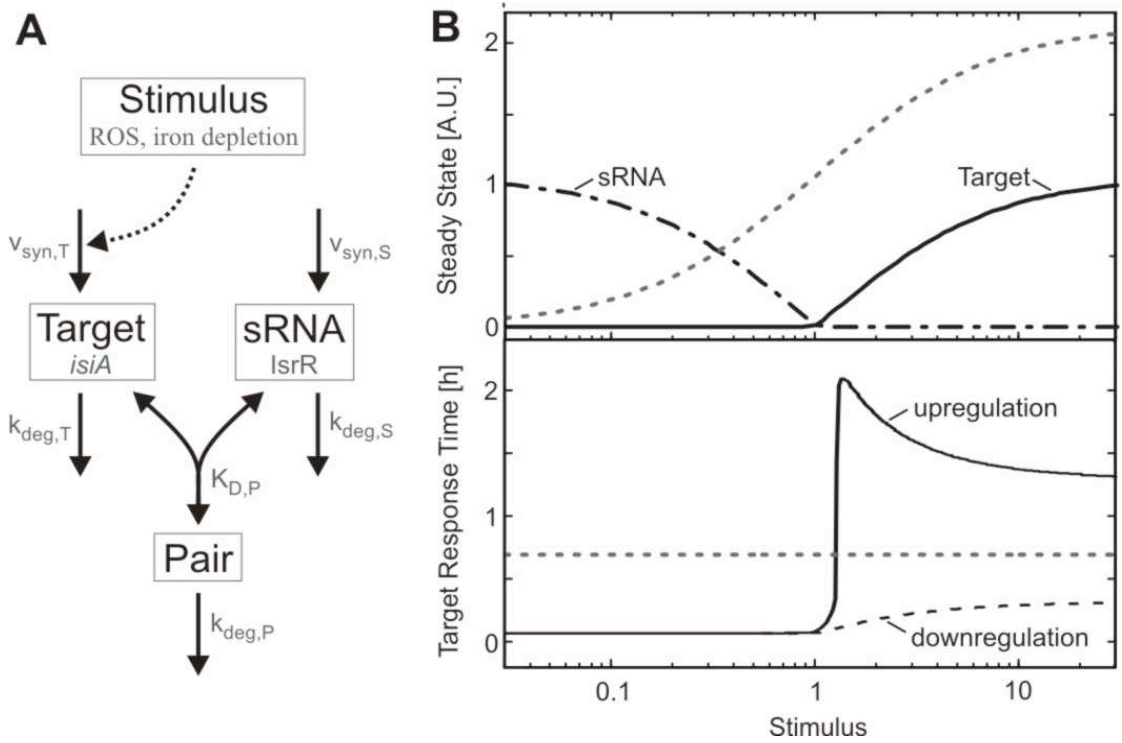

**Figure 2**: Experimental verification of simulated dynamical behaviour (A and B) and physiological relevance (C and D).

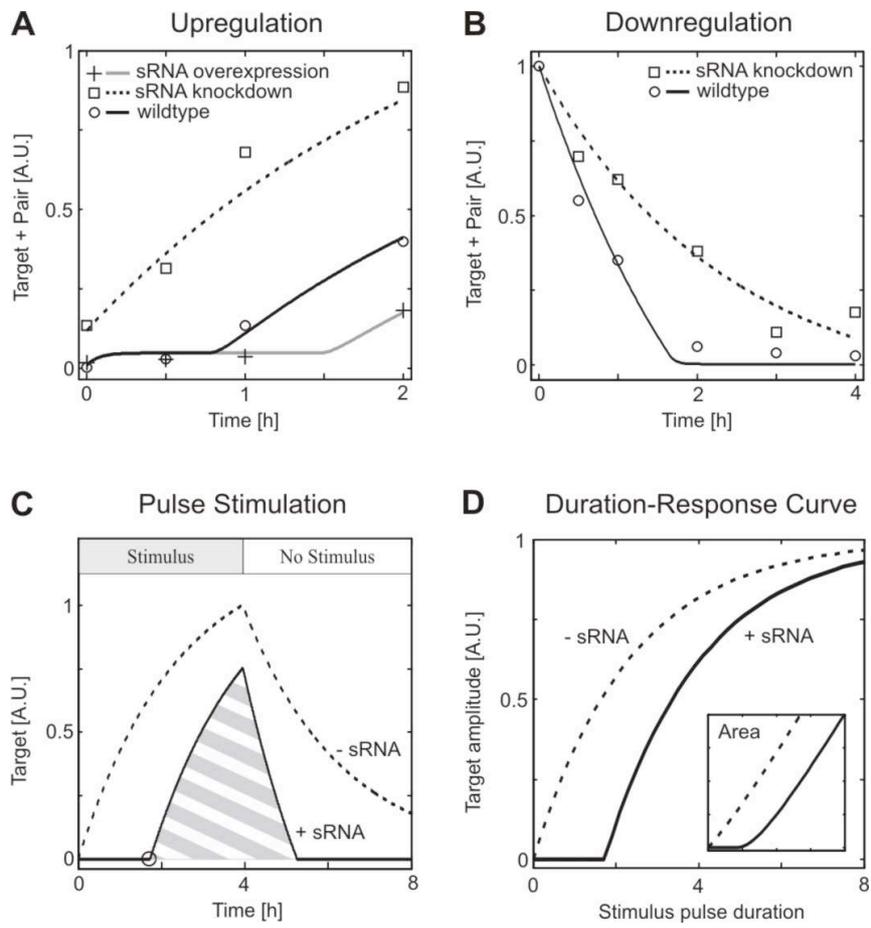